\documentclass[proof]{WileyASNA-v1}

\articletype{Article Type}%

\received{1 October 2020}
\revised{12 October 2020}
\accepted{25 October 2020}

\begin{document}

\title{One-parametric description for scalar field dark matter potentials}

\author[1,2]{Francisco X. Linares Cede\~no*}
\author[3]{L. Arturo Ure\~na-L\'opez**}

\authormark{Francisco X. Linares Cede\~no and L. Arturo Ure\~na-L\'opez}

\address[1]{\orgdiv{Instituto de F\'isica y Matem\'aticas}, \orgname{Universidad Michoacana de San Nicol\'as de Hidalgo, Edificio C-3, Ciudad Universitaria, CP. 58040}, \orgaddress{\state{Morelia, Michoac\'an}, \country{M\'exico}}}

\address[2]{\orgdiv{Mesoamerican Centre for Theoretical Physics}, \orgname{Universidad Autónoma de Chiapas, Carretera Zapata Km 4, Real del Bosque (Terán), 29040}, \orgaddress{\state{Tuxla Gutierrez, Chiapas}, \country{M\'exico}}}

\address[3]{\orgdiv{Departamento de F\'isica, DCI, Campus Le\'on}, \orgname{Universidad de
Guanajuato, 37150}, \orgaddress{\state{Le\'on, Guanajuato}, \country{M\'exico}}}

\corres{*\ \ \ linares@mctp.mx,\\ ** lurena@ugto.mx}

\abstract{We study the cosmological evolution for a scalar field dark matter model, by considering a parameterization of the evolution equations that allow us to unify in a single parameter a family of potentials: quadratic (free case), trigonometric (Axion-like case), and hyperbolic. After exploring the cosmological dynamics of this model, we perform a statistical analysis to study the viability of such model in comparison with the standard Cold Dark Matter model. We found that the free case is preferred over the other scalar field potentials, but in any case all of them are disfavored by the cosmological observations with respect to the standard model.  }

\keywords{Scalar field, Dark matter, Cosmology}

\maketitle

\section{Introduction}

Dark matter constitutes one of the open problems in modern cosmology, and in physics in general. Among the several proposals to address this problem, there are the \textit{Scalar Field Dark Matter} (SFDM) models. Initial contributions were made by considering the study of these models on galactic and cosmological scales \cite{Matos:1998vk,Guzman:1998vg,Matos:2000ng,Matos:2000ss,Matos:2000xi}, where numerical solutions were obtained for the linear density fluctuations, and it was observed that a cut-off scale due to the mass of the scalar field naturally arises. Such cut-off has important repercussions on the process of structure formation, as have been studied since then (for recent results see \cite{Diez-Tejedor:2014naa,Martinez-Medina:2014hka,Marsh:2015wka,Chen:2016unw,Gonzales-Morales:2016mkl,Marsh:2018zyw}).

On the other hand, the cosmological evolution of SFDM has been systematically studied in~\cite{Marsh:2010wq,Marsh:2012nm,Marsh:2013ywa,Hlozek:2014lca,Marsh:2015xka,Urena-Lopez:2015gur,Diez-Tejedor:2017ivd,Hlozek:2017zzf,Cedeno:2017sou,LinaresCedeno:2020dte,Urena-Lopez:2019kud}. Particularly, in \cite{Urena-Lopez:2015gur} the dynamics of SFDM with a quadratic potential was analyzed by rewriting the scalar field equations as a dynamical system and, by using an amended version of the Boltzmann code \textsc{class} \cite{Blas:2011rf,Lesgourgues:2011re}, the Cosmic Microwave Background (CMB) anisotropies as well as the matter power spectrum (MPS) were obtained. Later, in \cite{Cedeno:2017sou} the analysis was generalized by considering a Axion-like potential, i.e., a trigonometric cosine. Such potential is characteristic of QCD Axions \cite{Peccei:1977ur,Peccei:1977hh,Weinberg:1977ma,Wilczek:1977pj} and Axion-like particles \cite{witten1984some,Svrcek:2006yi,cicoli2012type,cicoli2014global,ringwald2014searching,ahn2016qcd}. When considering a trigonometric potential for the SFDM, a bump in the MPS at small scales (large $k$'s) is obtained. We shall mention that, whereas the fiducial mass is given by $m_{\phi}\simeq 10^{-22}$eV, Lyman-$\alpha$ observations seems to indicate that $m_{\phi}\geq 4\times 10^{-21}$eV \cite{Irsic:2017yje,Armengaud:2017nkf,Kobayashi:2017jcf}. Nonetheless, such constraint have been obtained for a SFDM with quadratic potential, and the bump induced by the trigonometric potential helps to alleviate this tension in the value of the scalar field mass, as was reported by \cite{Cedeno:2017sou}, and later by \cite{Leong:2018opi}.

The main goal of this paper, is to extend to a more generalized SFDM evolution through the inclusion of a hyperbolic potential, as those explored by \cite{Sahni:1999qe,Sahni:1999gb,Matos:2000ng,Matos:2000ss,Matos:2004rs}, and more recently by one of the authors~\cite{Urena-Lopez:2019xri}. As we will see, a proper change of variables will allow us to write the scalar field potential as
\begin{eqnarray}
     \label{potential}
     V(\phi) = \left\{
	       \begin{array}{ll}
		 m^2_{\phi}f^2\left[ 1 + \cos\left( \phi/f \right) \right]\, ,       &\mathrm{if\ } \lambda > 0 \\
		 \frac{1}{2}m^2_{\phi}\phi^2\, , &\mathrm{if\ } \lambda = 0 \\
		 m^2_{\phi}f^2\left[ 1 + \cosh\left( \phi/f \right) \right]\, ,       &\mathrm{if\ } \lambda < 0
	       \end{array}
	     \right.
   \end{eqnarray}
where $m_{\phi}$ is the mass of the scalar field and $f$ the so-called decay constant. As we shall see below, we will be able to compare the different predictions of the different families of SFDM potentials by changing the value of a single parameter, $\lambda$, which is associated to the decay constant $f$, as we will show later.

The structure of this work is as follows. In Sections~\ref{sfe} and~\ref{obs} we describe the background and linear perturbations equations for the SFDM, respectively. In Section~\ref{be} we calculate the Bayesian evidence of the SFDM models with respect to the standard Cold Dark Matter (CDM) model, to quantify the preference of cosmological observations for one model or another. Finally, in Section~\ref{conc} we discuss the main results and future perspectives of this work.

\section{Background evolution}
\label{sfe}

Let us consider a spatially-flat Friedmann-Robertson-Walker (FRW) line element,
\begin{equation}
ds^2 = -dt^2 + a^2(t)\left[ dr^2 + r^2\left(d\theta^2 + \sin^2\theta d\phi^2\right) \right]\, ,
\label{flrwsf}
\end{equation}
where $a(t)$ is the scale factor. The background equations for ordinary matter, as well as for SFDM $\phi$ endowed with a potential $V(\phi)$ are given by
\begin{subequations}
\label{eq:2}
  \begin{eqnarray}
    H^2 &=& \frac{\kappa^2}{3} \left( \sum_j \rho_j +
      \rho_\phi \right) \, , \\
    \dot{H} &=& - \frac{\kappa^2}{2} \left[ \sum_j (\rho_j +
      p_j ) + (\rho_\phi + p_\phi) \right] \, , \label{eq:2a} \\
    \dot{\rho}_j &=& - 3 H (\rho_j + p_j ) \,
    , \quad \quad
    \ddot{\phi} = -3 H \dot{\phi} - \frac{dV(\phi)}{d\phi}  \, , \label{eq:2b}
  \end{eqnarray}
\end{subequations}
where $\kappa^2 = 8\pi G$. The dot denotes derivative with respect to cosmic time $t$, and $H=\dot{a}/a$ is the Hubble parameter. By introducing the following change of variables~\cite{Copeland:1997et,Urena-Lopez:2015odd,Urena-Lopez:2015gur,Cedeno:2017sou,Roy:2018nce,LinaresCedeno:2020dte},
\begin{subequations}
\begin{eqnarray}
&&\Omega^{1/2}_\phi \sin(\theta/2) =  \frac{\kappa \dot{\phi}}{\sqrt{6} H} \, ,\quad \Omega^{1/2}_\phi \cos(\theta/2) = \frac{\kappa V^{1/2}}{\sqrt{3} H} \, ,\\
&& y_1 \equiv -\frac{2\ \sqrt[]{2}}{H}
   \partial_{\phi}V^{1/2} \, ,\quad y_2 \equiv -\frac{4\ \sqrt[]{3}}{H\kappa}\partial^{2}_{\phi}V^{1/2}\, ,
\label{dynsyscart}
\end{eqnarray}
\end{subequations}
it can be shown that the Klein-Gordon equation is written as
\begin{subequations}
\label{eq:new4}
  \begin{eqnarray}
  \theta^\prime &=& -3 \sin \theta + y_1 \, , \label{eq:new4a} \\
  \Omega^\prime_\phi &=& 3 (w_{tot} - w_\phi)
  \Omega_\phi \label{eq:new4c} \, ,\\
  y^\prime_1 &=& \frac{3}{2}\left( 1 + w_{tot} \right) y_1 + \Omega_{\phi}^{1/2} \sin (\theta/2)y_2\, , \label{eq:new4b} 
\end{eqnarray}
\end{subequations}
where a prime indicates derivatives with respect to the e-fold number $N=\ln(a/a_i)$. 

Until this point, the dynamical equations are valid for any dark matter potential. Once the potential is specified, it can be shown that $y_2$ can be written in terms of the other variables, i.e., $y_2 = y_2(\theta\, ,\Omega_{\phi}\, , y_1)$. We will be interested in trigonometric and hyperbolic functions of the scalar field, whereas the quadratic potencial will be a particular case. For the foregoing potentials we define the parameter $\lambda \equiv \mp 3/\kappa^2f^2$, and then the variable $y_2$ is written as:
\begin{equation}
    y_2 = \lambda \Omega^{1/2}_{\phi}\cos(\theta/2)\, .
\end{equation}
In this way, when we consider positive values of $\lambda$ we will be dealing with a scalar field endowed with a trigonometric cosine potential, and for $\lambda<0$ the potential will be the hyperbolic cosine. Notice that $\lambda = 0$ corresponds to the limit $f\rightarrow \infty$, for which we recover the standard Fuzzy Dark Matter scenario where the SFDM potential is a quadratic function $V\propto \phi^2$. 

\begin{center}
\begin{figure}[htp!]
    \centering
   \includegraphics[width=1.0\linewidth]{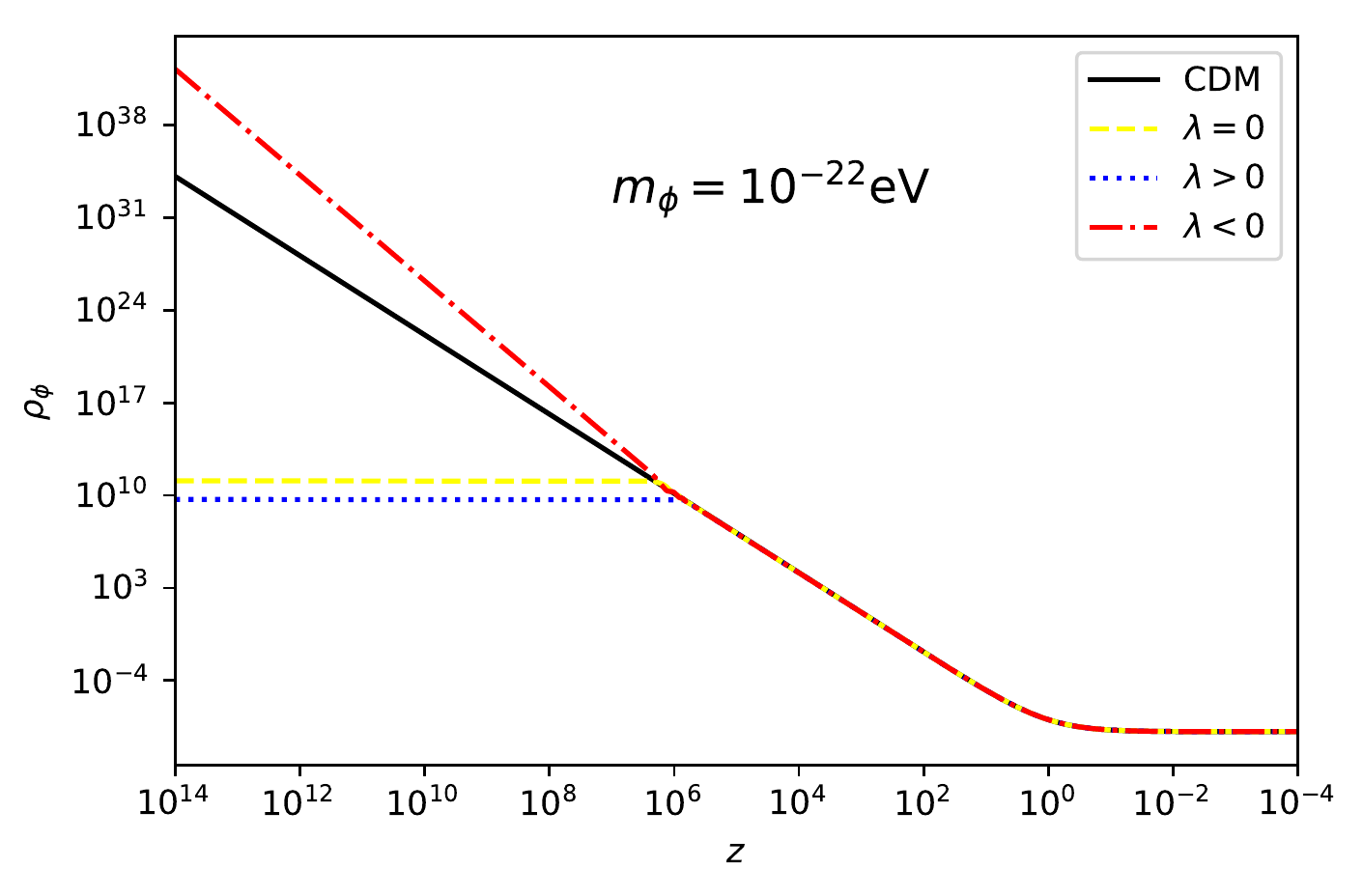}
    \caption{Scalar field energy density $\rho_{\phi}$ as function of the redshift $z$ for the quadratic (yellow line), trigonometric (blue line), and hyperbolic potential (red line). The evolution of CDM in the case of the standard $\Lambda$CDM model is shown for comparison (black line).}
    \label{rho}
\end{figure}
\end{center}

To study the behaviour of the cosmological evolution for both background and linear perturbations for the potentials~\eqref{potential}, we choose fixed values for the potential parameter given by $\lambda = \left\lbrace 0, 10^5, -3.225\times 10^3\right\rbrace$ for the quadratic ($\lambda=0$), trigonometric ($\lambda>0$), and hyperbolic ($\lambda<0$) potential, respectively. Likewise, all the numerical solutions we show in what follows were obtained for the fiducial mass $m_{\phi}=10^{-22}$eV. The cosmological evolution of the SFDM energy density $\rho_{\phi}$ is shown in Figure~\ref{rho}, where it can be seen that it is until some moment of the cosmological evolution that $\rho_{\phi}$ eventually evolves as CDM (black) for the three cases, quadratic (yellow), trigonometric (blue), and hyperbolic (red) potential.

The characteristic delay in the CDM--like evolution of $\rho_{\phi}$ due to the nonlinearities of the trigonometric potential can be appreciated in Figure~\ref{rho}, as was reported in detail in~\cite{Cedeno:2017sou,LinaresCedeno:2020dte}. Whereas $\rho_{\phi}$ for the quadratic and trigonometric potentials remains constant at early times, in the hyperbolic case the energy density evolves as a radiation component. 

It must be noticed that the SFDM equation of state (EoS), given by $w_\phi = -\cos \theta$, oscillates very rapidly in the range $[-1:1]$, and these oscillations can be averaged so that the effective EoS on cosmological times is $\langle w_\phi \rangle = 0$. This averaging procedure is done numerically as explained in~\cite{Urena-Lopez:2015gur}, which does not alter the true evolution of the SFDM energy density (see also~\cite{Cookmeyer:2019rna} for a comparison between cut-off procedures used in the literature.).

\section{Linear density perturbations}\label{obs}

When considering linear perturbations, the line element in the synchronous gauge is given by
\begin{equation}
ds^2 = -dt^2+a^2(t)(\delta_{ij}+h_{ij})dx^idx^j\, ,\ \phi(\vec{x},t) = \phi(t)+\varphi(\vec{x},t)\, ,
\label{mfpert}
\end{equation}
where $h_{ij}$ and $\varphi$ are the metric and scalar field perturbations, respectively. The linearized Klein-Gordon equation for the scalar field perturbation is written (in Fourier space) as \cite{Ratra:1990me,Ferreira:1997au,Ferreira:1997hj,Perrotta:1998vf}
\begin{equation}
  \ddot{\varphi}(\vec{k},t) = - 3H \dot{\varphi}(\vec{k},t) - \left[\frac{k^2}{a^2} + \frac{\partial^2 V(\phi)}{\partial \phi^2}\right] \varphi(\vec{k},t) -
  \frac{1}{2} \dot{\phi} \dot{\bar{h}} \, . \label{eq:13}
\end{equation}

In a similar way to the procedure we have done for the Klein-Gordon equation~\eqref{eq:2b}, we propose the following change of variables for the scalar field perturbation $\varphi$ and its derivative $\dot{\varphi}$ \cite{Urena-Lopez:2015gur,Cedeno:2017sou} 
\begin{equation}
-\Omega^{1/2}_{\phi}e^{\alpha}\cos(\vartheta/2) = \sqrt{\frac{2}{3}} \frac{\kappa \dot{\varphi}}{H}\, , \label{eq:22a} \quad
    -\Omega^{1/2}_{\phi}e^{\alpha}\sin(\vartheta/2) = \frac{\kappa y_1 \varphi}{\sqrt{6}}\, .
\end{equation}
If we introduce the following transformation
\begin{equation}
\delta_0=-e^{\alpha}\sin[(\theta - \vartheta)/2]\, , \quad \delta_1=-e^{\alpha}\cos[(\theta - \vartheta)/2]\, ,
\label{d0d1}
\end{equation}
we obtain
\begin{subequations}
\label{eqnewdeltas}
\begin{eqnarray}
\delta^\prime_0 &=&  \left[-3\sin\theta-\frac{k^2}{k^2_J}(1 - \cos \theta) \right] \delta_1 + \frac{k^2}{k^2_J} \sin \theta \delta_0 \nonumber \\
&&- \frac{\bar{h}^\prime}{2}(1-\cos\theta) \, , \label{d0} \\
\delta^\prime_1 &=& \left[-3\cos \theta - \frac{k_{eff}^2}{k_J^2} \sin\theta \right] \delta_1 + \frac{k_{eff}^2}{k_J^2} \left(1 + \cos \theta \right) \, \delta_0 \nonumber \\
&& - \frac{\bar{h}^\prime}{2} \sin \theta \, ,  
\label{d1}
\end{eqnarray}
\end{subequations}
where we have defined the \textit{effective Jeans wavenumber} as
\begin{equation}
    k^2_{eff} \equiv k^2 - \lambda a^2 H^2 \Omega_\phi/2\, .
    \label{keff}
\end{equation}

The expression~\eqref{keff} encodes the effects that the SFDM potential can have on the evolution of linear density perturbations. The system of equations~\eqref{eq:new4} and~\eqref{eqnewdeltas} have been studied for the cases $\lambda= 0$ (quadratic potential) \cite{Urena-Lopez:2015gur} and $\lambda >0$ (trigonometric potential) \cite{Cedeno:2017sou}. In the former, the mass of the scalar field $m_{\phi}$ induces a cut-off scale in the MPS at small scales. In the latter, the non-linear effects of the trigonometric potential induce a cut-off as well, but with a bump at such small scales. The case $\lambda < 0$ (cosh potential) has been studied in~\cite{Matos:2000ng,Urena-Lopez:2019xri}, and the resultant MPS is the same as in the case of the quadratic potential for the same value of the SFDM mass $m_\phi$.

The numerical solutions obtained from the linear perturbation equations allow us to build cosmological observables, such as the temperature anisotropies of the Cosmic Microwave Background (CMB), and the aforementioned MPS. For this we use the Boltzmann code \textsc{class}~\cite{Blas:2011rf,Lesgourgues:2011re}, which was amended to solve the SFDM equations of motion together with all other relevant cosmological equations. More details about the amendments can be found in~\cite{Cedeno:2017sou,LinaresCedeno:2020dte,Urena-Lopez:2015gur}.

\begin{center}
\begin{figure}[htp!]
    \centering
    \includegraphics[width=\linewidth]{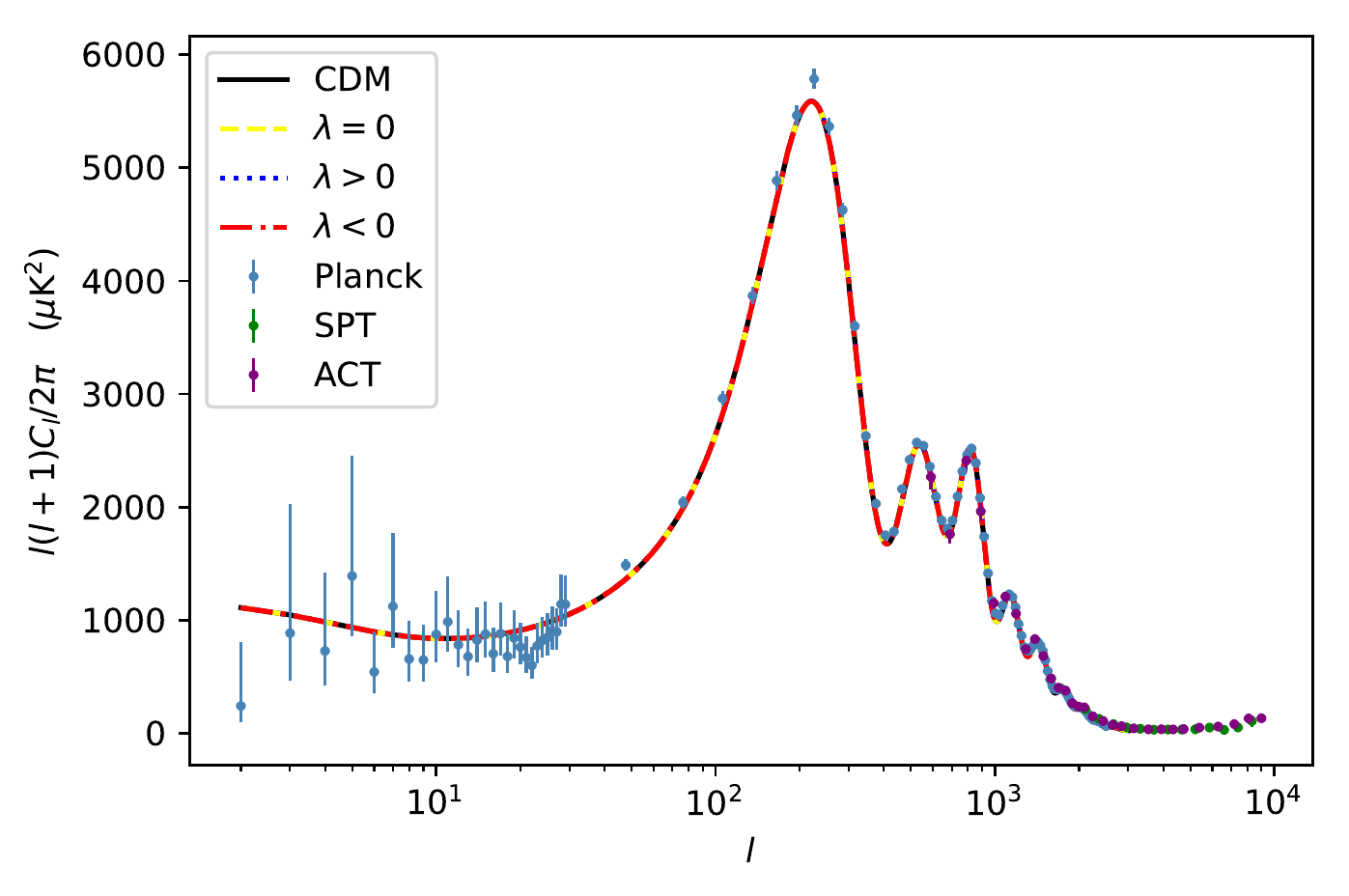}
    \includegraphics[width=\linewidth]{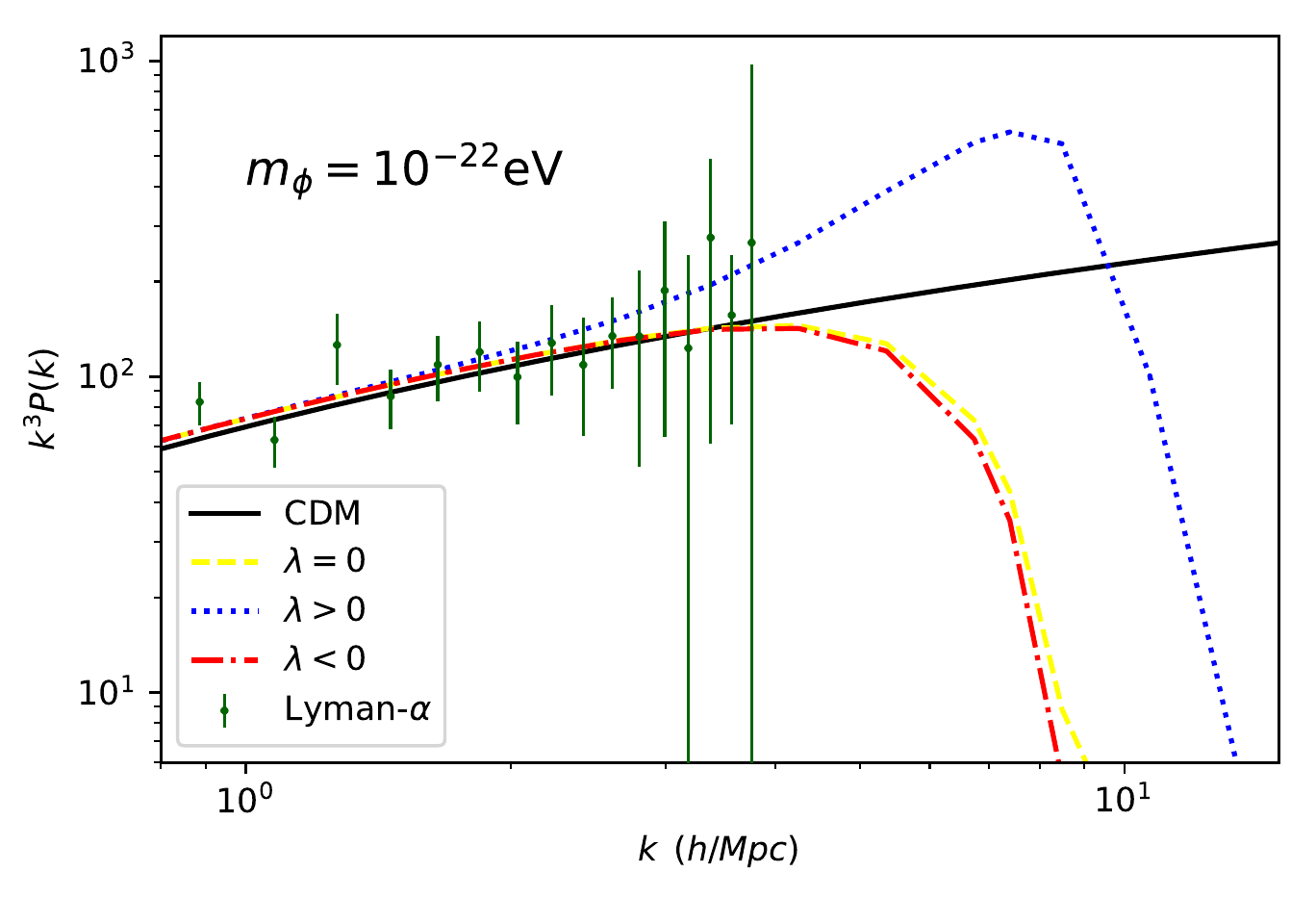}
    \caption{(Top) Temperature anisotropies of the CMB. For a scalar field mass given by $m_{\phi}=10^{-22}$eV, the differences between SFDM and CDM are negligible. The data points are from the Planck, SPT and ACT collaborations. (Bottom) MPS for each of the scalar field dark matter models. Whereas for the trigonometric potential there is a bump at large wavenumbers besides the characteristic cut--off, the quadratic and hyperbolic potentials are very similar in their prediction of the structure formation. The data points are from~\cite{Chabanier:2018rga} and are shown for reference.}
    \label{cmb}
\end{figure}
\end{center}

We show in the top panel of Figure~\ref{cmb} the CMB temperature for each of the SFDM models, as well as that for the standard CDM model. It can be seen that there are not notorious differences in the anisotropies spectra between the cosmological models. Likewise, the MPS is shown in bottom panel of Figure~\ref{cmb}. The characteristic cut--off at large wavenumbers is present for all the potentials. However, for the trigonometric case there is also a bump at the cut--off scale, which have been already analyzed in~\cite{Cedeno:2017sou}. Notice that the MPS, as is also the case for the CMB spectrum in Figure~\ref{cmb}, for the quadratic and hyperbolic potentials are very similar, which indicates the existence of a degeneration between these two models.

\section{Bayesian evidence}
\label{be}
The standard procedure in cosmology for the comparison of theoretical models with data is to use the Bayes theorem and perform large numerical Monte Carlo runs to calculate the posterior distributions of the model parameters, with this one can estimate the values of the latter that correspond to the maximum likelihood. This kind of studies can be found in~\cite{Cedeno:2017sou,Hlozek:2014lca} for the quadratic and trigonometric potentials using available cosmological data. Such study does not exist yet for the cosh potential, but given its similarities with the quadratic one, one can safely assume that the resultant constraints in the parameteres will be equally similar. For brevity, we will follow here a different approach and calculate the so-called Bayesian evidence for each one of the models, which will allow us to compare the models one with each other and see whether available data show any preference for any of them. 

Bayes theorem states that, given some observed data $D$, the probability of a model $M$ described in terms of a set of parameters $\Theta$, is given by the \textit{Posterior} $\mathcal{P}$: 
\begin{equation}
    \mathcal{P}(\Theta \mid D,M)= \frac{\mathcal{L}(D\mid \Theta,M)\Pi(\Theta\mid M)}{\mathcal{E}(D\mid M)},
    \label{eq:bayes}
\end{equation}
where $\mathcal{L}$ is the Likelihood function, $\Pi$ represents the set of priors, which contain the \textit{a priori} information about the parameters of the model, and $\mathcal{E}$ is the so-called Evidence, to which we pay particular attention\footnote{For a comprehensive review of Bayesian model selection, we refer the reader to \cite{Trotta:2008qt}.}.

The evidence $\mathcal{E}$ normalises the area under the posterior $\mathcal{P}$, and is given by
\begin{equation}
  \mathcal{E}(D\mid M)= \int d\Theta \mathcal{L}(D|\Theta,M)\Pi(\Theta|M)\, . 
  \label{eq:evidence}
\end{equation}
When comparing two different models $M_1$ and $M_2$ using the Bayes' theorem \eqref{eq:bayes}, the ratio of posteriors of the two models $\mathcal{P}_1$ and $\mathcal{P}_2$ will be proportional to the ratio of their evidences. This leads to the definition of the \textit{Bayes Factor} $B_{12}$, which in logarithmic scale is written as
\begin{equation}
    \log B_{12} \equiv \log \left[\mathcal{E}_1(D\mid M_1)\right] - \log \left[\mathcal{E}_2(D\mid M_2)\right] 
    \label{BF}
\end{equation}
If $\log B_{12}$ is larger (smaller) than unity, the data favours model $M_1$ ($M_2$). To assess the strength of the evidence contained in the data,~\cite{Jeffreys1961} introduces an empirical scale, which quantifies the strength of evidence for a corresponding range of the Bayes factor. We follow the convention of~\cite{Kass:1995loi,10.2307/2337598} in presenting a factor of two with the natural logarithm of the Bayes factor.

To calculate the evidence, we first generate a series of MCMC with the cosmological estimator parameters \textsc{monte python}, developed by~\cite{Audren:2012wb}, and using data from Planck 2018 Collaboration, see~\cite{Aghanim:2018eyx,Aghanim:2019ame}\footnote{Based on observations obtained with Planck (http://www.esa.int/Planck), an ESA science mission with instruments and contributions directly funded by ESA Member States, NASA, and Canada.}. We then follow a recent proposal by~\cite{Heavens:2017afc}, in that the unnormalized posterior $\tilde{\mathcal{P}}(\Theta\mid D,M)$ is proportional to the number density $n(\Theta\mid D,M)$. Since the number density is given by
\begin{equation}
    n(\Theta\mid D,M) = N\ \mathcal{P}(\Theta\mid D,M) = N\ \frac{\tilde{\mathcal{P}}(\Theta\mid D,M)}{\mathcal{E}(D \mid M)},
\end{equation}
where $N$ is the lenght of the chain, then the evidence is given by
\begin{equation}
    \quad \mathcal{E}(D \mid M) = a\ N\, .
\end{equation}
By determining the proportionality constant $a$, it is possible to calculate the evidence $\mathcal{E}$ directly from the MCMC chains. The software developed for this approach is called \textsc{MCEvidence}, and it has been used to calculate the evidence for several alternative cosmological models, as it is shown by~\cite{Heavens:2017hkr,Kouwn:2017qet,Binnie:2019dwt,Gomez-Valent:2020mqn,Archidiacono:2020dvx}.

\begin{center}
\begin{figure}[htp!]
    \centering
   \includegraphics[width=\linewidth]{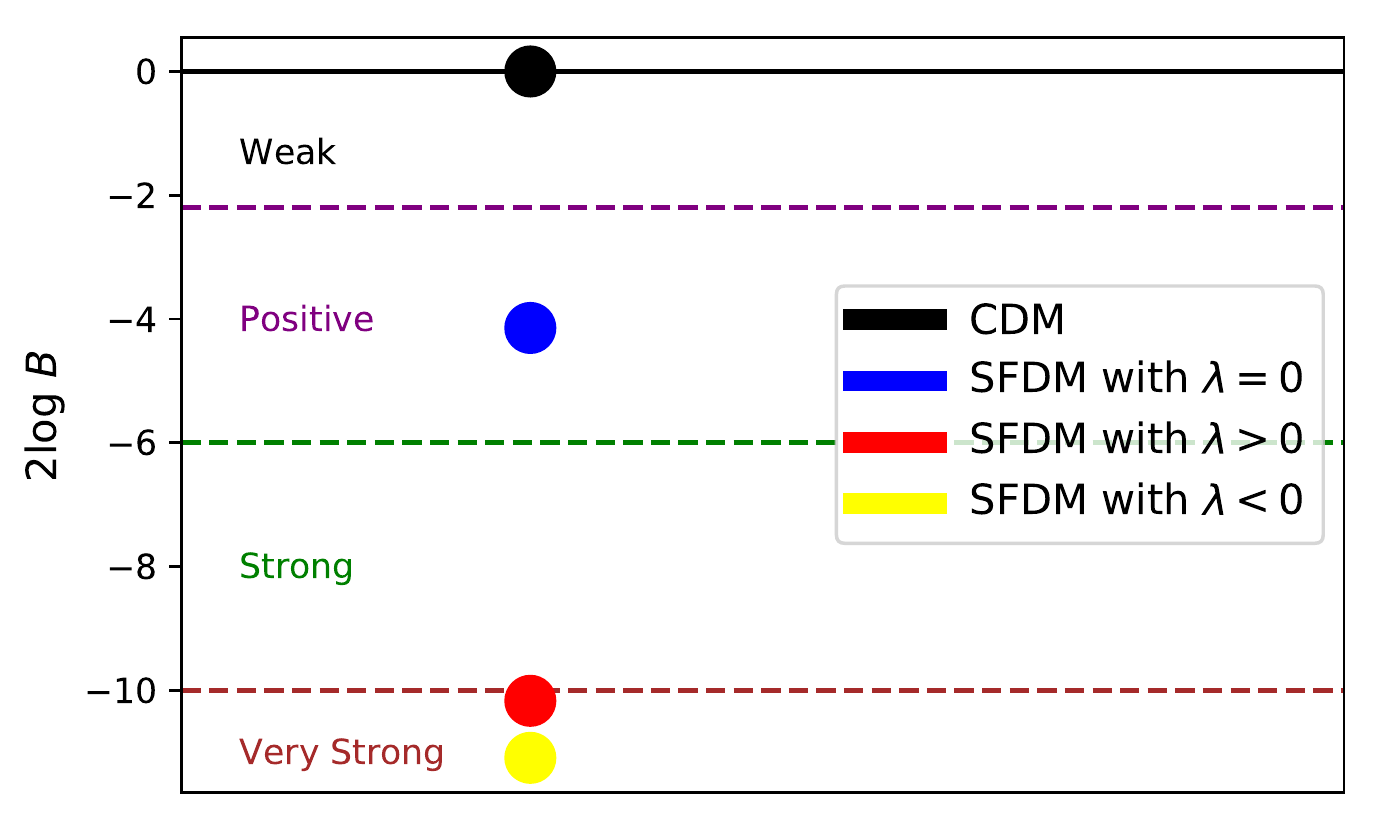}
    \caption{Bayes factor~\eqref{BF} estimated for the SFDM models discussed in the text with respect to the standard CDM one. See the text for more details.}
    \label{evid}
\end{figure}
\end{center}

Figure~\ref{evid} shows the Bayes factor comparing all the SFDM models against the standard CDM. It can be seen that the CDM model is preferred over the SFDM models. The horizontal lines indicate the Jeffrey’s scale, from which we obtain that the SFDM with quadratic potential is disfavored, with positive evidence, against the CDM model, and that the SFDM with trigonometric and hyperbolic are also disfavored, but now with strong and very strong evidence, respectively.

\section{Final remarks}
\label{conc}
In this paper we have studied three different SFDM models, taking advantage of a formalism that allows the calculation of the cosmological observables in a unified manner. The SFDM models have been considered in the literature as serious alternatives to the standard CDM one, being all of them consistent with current observations at hand when considered individually. Here we present the first comparison of them with respect to the CDM case, using the Bayesian evidence.

Our preliminary results indicate a preference of the cosmological observations for the CDM model, with some positive evidence against the free case ($\lambda =0$), and strong and very strong evidence against the trigonometric ($\lambda <0$) and cosh ($\lambda >0$) potentials. Part of the explanation is that SFDM models introduce at least one extra parameter, which may be penalized by the Bayesian evidence. However, the strong reason behind the rejection of the SFDM models can be that current cosmological observations do not yet indicate the existence of characteristic scales related to the dark matter component, e.g., the clear presence of a cut-off in the MPS of density perturbations. On the other hand, we have explored the full possibility of varying both, the scalar field mass and the potential parameter $\lambda$. This could lead to combinations in the parameter space that are ruled out by CMB observations. 

As future work, we will analyze these models for fixed masses, and focusing on the potential parameter, in order to establish a more proper comparison against CDM. Besides, it will be useful to include other set of data, such as those from Lyman-$\alpha$ forest, to be able to constraint the SFDM models by considering the cut-off at large wavenumbers in the MPS, which is a clear distinction of these models in comparison with the CDM case. \\

\textbf{Acknowledgments.-} Francisco X. Linares Cedeño acknowledges the receipt of the grant from the Abdus Salam International Centre for Theoretical Physics, Trieste, Italy; and CONACYT and the Programa para el Desarrollo Profesional Docente for financial support.  This work was partially supported by Programa para el Desarrollo Profesional Docente; Direcci\'on de Apoyo a la Investigaci\'on y al Posgrado, Universidad de Guanajuato under Grant No. 099/2020; CONACyT M\'exico under Grants No. A1-S-17899, 286897, 297771, 304001; and the Instituto Avanzado de Cosmología collaboration.

\bibliography{IWARA20}%

\end{document}